\documentclass[aps,prl,10pt,preprint,nofootinbib]{revtex4-1}
\usepackage{graphicx}
\usepackage{amsmath}
\usepackage{amssymb}
\usepackage{color}

\begin{document}

\title{Determination of Electron Band Structure using Temporal Interferometry}

\author{Liang Li$^{1}$, Pengfei Lan$^{1}$}
\email{pengfeilan@hust.edu.cn}
\author{Lixin He$^{1}$, Wei Cao$^{1}$, Qingbin Zhang$^{1}$}
\author{Peixiang Lu$^{1,2}$}
\email{lupeixiang@hust.edu.cn}

\affiliation{%
 $^1$Wuhan National Laboratory for Optoelectronics and School of Physics, Huazhong University of Science and Technology, Wuhan 430074, China\\
$^2$Hubei Key Laboratory of Optical Information and Pattern Recognition, Wuhan Institute
of Technology, Wuhan 430205, China
}%

\date{\today}

\begin{abstract}
We propose an all optical method to directly reconstruct the band structure of semiconductors. Our scheme is based on the temporal Young's interferometer realized by high harmonic generation (HHG) with a few-cycle laser pulse. As a time-energy domain interferometric device, temporal interferometer encodes the band structure into the fringe in the energy domain. The relation between the band structure and the emitted harmonic frequencies is established. This enables us to retrieve the band structure from the HHG spectrum with a single-shot measurement. Our scheme paves the way to study matters under ambient conditions and to track the ultrafast modification of band structures.

\end{abstract}
\maketitle

The electron band structure can be understood as the material fingerprint in the reciprocal space and determines many properties of materials. Measuring the band structure is of great importance for understanding the properties of materials. Usually, the electron band structure of solids is mapped by independently measuring the momentum and the energy of incoherent electrons by the angle-resolved photoemission spectroscopy (ARPES) \cite{ARPES}. Normally, only the sample area close to the surface can be investigated, because the electron mean free paths in solids are typically in the angstrom scale. In addition, the photoelectrons are sometimes difficult or impossible to be detected because of the scattering by ambient conditions. These limit the access to the band structure by ARPES from bulk matters or under ambient conditions \cite{Shuvaev2017}.

In recent years, high harmonic generation (HHG) has been observed experimentally from a wide variety of solid media \cite{Ghimire2010,Vampa2015,Huber2015,You2017,Tanaka2017,Werner2018,Rashid2018}, from general semiconductors to novel materials such as graphene. This opens avenues toward attosecond science on the platform of solid-state materials \cite{Krausz2012,Schultze2012,Zaks2012,Langer2016,Langer2018}, and suggests new approaches for crystallographic analysis and probing the electronic properties of solids \cite{Hoegen2018,Worner2018,Banks2017,Chao2018,Wang2016}. Different from the traditional electronic techniques, the method based on HHG measures the high harmonic photons. Thus it is an all-optical approach and can be used to bulk materials and ambient conditions. Moreover, it has high temporal resolution, making it potentially to study the ultrafast transient modifications of band structures. Very recently, several works \cite{Luu,Vampa1,Lanin} proposed methods to measure the band structure of ZnO and ZnSe by HHG. In their methods, the harmonic spectra are detected as a function of the field intensity or the time delay between the two-color field. Then, the band structure is retrieved by comparing the measured harmonic yields with those calculated from a set of trial bands, and find the one that best fits the experiments. These methods however both rely on the calculations of high harmonic spectra with numerical models and the reconstructions are partially based on the theoretical simulations or assumptions. The band reconstruction is very time-consuming. It sometimes becomes even computationally prohibitive to get a convergent retrieval when the number of undetermined parameters become large. Moreover, these methods need multi-shot measurements to obtain the intensity (or delay) dependent high harmonic yields. It prevents from effectively capturing the band structure and realizing the time-resolved measurements of ultrafast dynamics.

\begin{figure}[!t]
	\includegraphics[width=7.6cm]{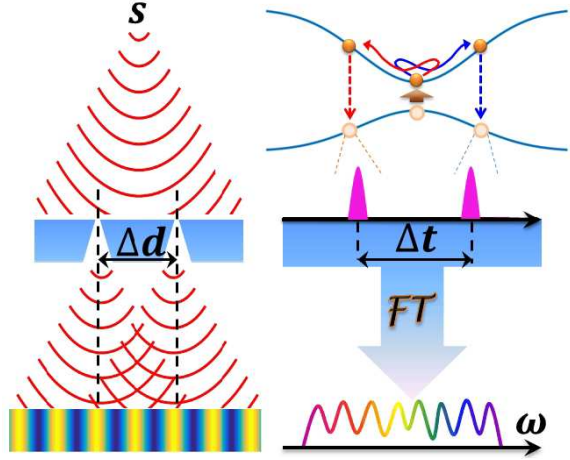}
	\caption{The sketch of the two-slit interferometer from HHG in solids. In HHG, an electron source is generated at the top of the valence band and two bursts are emitted at the two adjacent half laser cycles. These two bursts constructs a two-slit interferometer and interference fringes will be observed in the frequency domain.}\label{fig1}
\end{figure}

In this Letter, we demonstrate a method to directly reconstruct the band structure. Our scheme is based on the generalized temporal Young's interferometer \cite{Wooters1979,Scully1991,YouYS2017}. As illustrated in Fig. \ref{fig1}, we construct a temporal two-slit with a few cycle laser field. The HHG are dominantly contributed by two emissions, because the ionization are constrained within an optical cycle near the peak of the envelope. Two representative trajectories (red and blue lines in Fig. 1) propagate oppositely at two sequent half cycles accompanied with harmonics emission at different times (the middle line of Fig. 1). The interference fringe comes from this two emissions, i.e., the peaks of HHG spectrum, is very sensitive to the phase different in the wave front between the two slits ($\Delta S$ in Eq. \ref{peaks} shown hereinafter), which is encoded by the band structure. Then, one can directly retrieve the band structure by monitoring the interference fringe.

To show how one can retrieve the band structure from the HHG signal, we first explain the concept of temporal interferometry. We start from the time-dependent Schr\"{o}dinger equation (TDSE) \cite{Becker2019} for a general solid system, 
\begin{align}\label{TDSE1}
	[i\frac{\partial}{\partial t}-\hat{H}_{\text{free}}-\hat{H}_{\text{int}}]|\Psi_{c,\textbf{k}_{0}}\rangle=\hat{H}_{\text{int}}|\Psi_{v,\textbf{k}_{0}}\rangle=\sigma
\end{align}
where $\hat{H}_{\text{free}}$ is the field free Hamiltonian and the $\hat{H}_{\text{int}}$ is the interaction Hamiltonian. The wavefunction is decomposed into the ground state $|\Psi_{v}\rangle$ of the Hamiltonian $\hat{H}_{\text{free}}$ and the remainder $|\Psi_{c}\rangle$, which includes the conduction states. The term $\sigma$ can be thought of as a source of electrons in ground states. Within this form, the high harmonic generation process can be expressed by a series emissions in time domain \cite{Li2019,SM}
\begin{align}\label{HHG}
P(\Omega,t) &= \sum_{t',t_{r},\textbf{k}_{0}}^{mn}a_{mn}(t',t_{r},\textbf{k}_{0})e^{-iS_{mn}(t',t_{r},\textbf{k}_{0})}\textbf{p}_{mn}(t',t_{r},\textbf{k}_{0})
\end{align}
The parameter $t',t_{r},\textbf{k}_{0}$ satisfied the saddle point equation. $a(t',t_{r},\textbf{k}_{0})$ is the weight of the channel and $\textbf{p}_{mn}(t',t_{r},\textbf{k}_{0})$ is the polarization between the electron-hole pairs when recollision occurs. Here the intraband terms are not shown, because with the parameters we used in this paper, the harmonic yield beyond the band gap is mainly contributed by the interband terms. Note that the above derivation is for a general solid system. According to Eq. \ref{HHG}, one can retrieved the band structure by measuring the interference fringe as long as the signal from the target bands can be identified.
Typical semiconductors exhibit band structure with multiple valence and conduction bands. However, in many cases, coupling between multiple bands is expected to be negligible for wide separated bands, or even closely spaced bands of similar symmetry. As shown in previous works \cite{Vampa1,Lanin,Wu2015,McDonald2015}, the signal of the target bands can be identified from the multi-valence and multi-conduction bands under appropriate laser parameters. Then, one can retrieve the band gap $E_{m}-E_{n}$ by using the temporal interferometry.

To demonstrate our scheme, we perform the simulated experiment with ZnO. The same band dispersion $E_{m}(\textbf{k}) = E_{m,x}(k_{x})+E_{m,y}(k_{y})+E_{m,z}(k_{z})$ as that in Ref. \cite{Vampa1,Vampa2014,Goano2007} is used. The orientation of reciprocal lattice is chosen so that $\hat{x}\parallel\Gamma-M$, $\hat{y}\parallel\Gamma-K$, and $\hat{z}\parallel\Gamma-A$ (optical axis). The laser field is propagating along the optical axis $\hat{z}$. We adopt a Gaussian envelope and the electric field can be expressed as
\begin{align}\label{one_half_cycle}
\textbf{F}(t) = \hat{x}F_{0}e^{-2ln2(t/\tau)^{2}}cos(\omega t+\phi)
\end{align}
$\omega$ and $\phi$ is the frequency and carrier-envelope phase (CEP) of the laser field, respectively. The wavelength and the intensity of the laser field are 3 $\mu m$ ($\omega$ = 0.015 a.u.) and $9\times10^{11}W/cm^{2}$ ($F_{0}$ = 0.005 a.u.), respectively. Atom units (a.u.) is applied throughout this paper unless stated. A few-cycle laser pulse with $\tau = 1.5T_{0}$ is used, where $T_{0}=\frac{2\pi}{\omega}$ is an optical cycle (o.c.). Due to the restriction of the few-cycle pulse, only two trajectories at the pulse peak are dominant, which construct a two-slit separated approximately by half optical cycle. The high harmonic spectra can be obtained from the semiconductor Bloch equation (SBE) \cite{Vampa2014,Koch2008,Luu2016}
\begin{align}\label{SBE}
&\frac{d\pi(\textbf{K},t)}{dt} = -\frac{\pi(\textbf{K},t)}{T_{2}}-i\xi(\textbf{K},t)[2n_{v}(\textbf{K},t)-1]e^{-iS(\textbf{K},t)} \nonumber \\
&\frac{dn_{v}(\textbf{K},t)}{dt} = -i\xi^{*}(\textbf{K},t)\pi(\textbf{K},t)e^{iS(\textbf{K},t)}+c.c. \nonumber \\
&\frac{d[n_{v}(\textbf{K},t)+n_{c}(\textbf{K},t)]}{dt} = 0
\end{align}
where $n_{m}$ is the band population. $S(\textbf{K},t) = \int_{t_{0}}^{t}[E_{c}(\textbf{K}+\textbf{A}(t'))-E_{v}(\textbf{K}+\textbf{A}(t'))]dt'$ is the classical action, $\xi(\textbf{K},t) = \textbf{F}(t)\cdot \textbf{d}(\textbf{K}+\textbf{A}(t))$ is the Rabi frequency, and $\textbf{d}(\textbf{k}) = \textbf{d}^{*}_{cv}(\textbf{k})=\textbf{d}_{vc}(\textbf{k})=\langle c,\textbf{k}|\hat{\textbf{r}}|v,\textbf{k}\rangle$ is the transition dipole moment with $|m,\textbf{k}\rangle$ representing the Bloch states. $\textbf{K}=\textbf{k}-\textbf{A}(t)$ is the shifted crystal momentum with the vector potential $\frac{d\textbf{A}(t)}{dt} = -\textbf{F}(t)$, and the first Brillouin zone is also shifted to $\bar{BZ}=BZ-\textbf{A}(t)$. $T_{2}$ is a dephasing-time term describing the coherence between the conduction and valence bands. We choose $T_2 = 2.5$fs in the simulation. The numerical value of dipole moment at $\Gamma$ point, $\textbf{d}_{cv} = (3.46,3.46,3.94)$, is applied and the k-dependence is neglected in our analysis. For computational convenience, we perform the two-dimensional (2D) calculations (i.e. $k_{z}=0$ in reciprocal space). The reciprocal space are discretized by a grid with $481\times481$ points. By numerically solving the SBE, one can obtain the intraband $\textbf{J}_{ra}$ and interband $\textbf{J}_{er}$ currents as follows,
\begin{align}\label{SBE_currents}
\textbf{J}_{ra}(t) &= \sum_{m=c,v}\int_{\bar{BZ}}\textbf{v}_{m}[\textbf{k}+\textbf{A}(t)]n_{m}(\textbf{K},t)d^{3}\textbf{K} \\
\textbf{J}_{er}(t) &= \frac{d}{dt}\int_{\bar{BZ}}\textbf{d}[\textbf{k}+\textbf{A}(t)]\pi(\textbf{K},t)e^{iS(\textbf{K},t)}d^{3}\textbf{K}+c.c.
\end{align}
where $\textbf{v}_{m}(\textbf{k}) = \triangledown_{\textbf{k}}E_{m}(\textbf{k})$ are the group velocity in band $m$. High harmonic spectra are obtained from the Fourier transform of these currents. With the parameters used in this work, the harmonic yields beyond the minimum band gap are dominated by the interband currents. Hereinafter, we only consider the interband harmonics.

\begin{figure}[!t]
	\includegraphics[width=8.6cm]{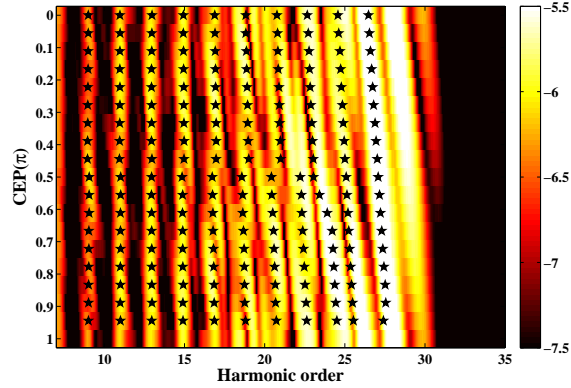}
	\caption{The harmonic spectra for different CEP. The stars mark the results from the two-slit interferometer.}\label{fig2}
\end{figure}

Figure \ref{fig2} shows the HHG spectra with the CEP changing from 0 to $\pi$. The harmonic order is identified by the frequency of the laser field, and the harmonics with photon energy lower than the minimum band gap are not shown. As shown in Fig. \ref{fig2}, the harmonic peaks are deviated from the odd harmonics ($O_{q} = (2q-1)\omega$), which are expected with a multi-cycle driven laser. For $\phi=0$, the harmonic peaks are red-shifted compared to the odd harmonics. With the increase of the harmonic order, the frequency shift becomes more and more evident. When the CEP changes from 0 to 0.5$\pi$, the harmonic peaks shift to higher photon energy, and even become blue-shifted compared to the odd harmonics. When the CEP changes from 0.5 to 1$\pi$, the harmonic peaks are shown similar structure to that from 0 to 0.5$\pi$.

\begin{figure}[!t]
	\includegraphics[width=8.6cm]{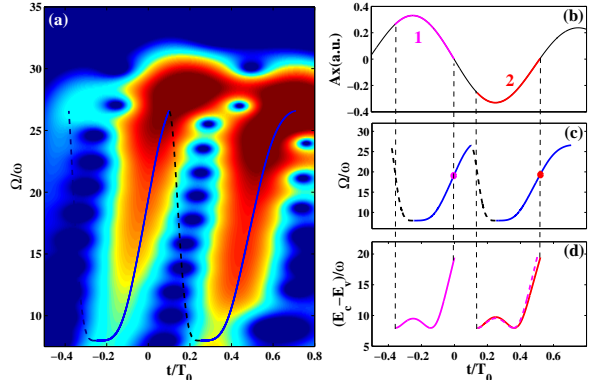}
	\caption{(a) Time-frequency spectrogram for $\phi=0$. (b) The vector potential of the electric field. ``1'' and ``2'' marks two typical channels for an identical photon energy. (c) The emission energy versus the ionization time and the emission time obtained with Eq. 8. (d) The paths of ``1'' and ``2'' in energy domain.}\label{fig3}
\end{figure}

To quantitatively understood the above results, we show the time-frequency spectrogram of HHG in Fig. \ref{fig3}(a). Only the signal near the pulse center is shown, because the signal outside this region is relatively weaker and will not influence the discussion. As shown in Fig. \ref{fig3}(a), there are two dominant emissions, which are contributed by the electrons ionized at adjacent half cycle at the center of the few-cycle pulse. We choose two representative emissions with the same energy, which are refered as ``1'' and ``2''  in Fig. \ref{fig3}(b). These two emission construct a temporal two-slit interferometer. The interference can be described by $d(t) = d(t-t_{1})+d(t-t_2)e^{i(\pi+\Delta S)}$. Here $\Delta t = t_{2}-t_{1}$ is the silts gap and $\Delta S = S_{2}-S_{1}$ is the phase difference between the channels ``1'' and ``2''. Note that $\pi$ phase shift is due to the reversal sign of the electric field. The slit gap and phase difference varies with the emission photon energy. As a time-energy domain interferometric device, fringes in energy domain are expressed as $I(\Omega) \propto |1+e^{i[\Omega*\Delta t+\Delta S+\pi]}|^2$. We have ignored the amplitude difference of these two emissions. Then we can obtain the constructive interference, i.e., the peak in the HHG spectrum,
\begin{align}\label{peaks}
\Omega = (\frac{T_{0}}{2\Delta t})O_{q}-\Delta S/\Delta t
\end{align}
The displacement of interference fringe, that is, the frequency shift between the harmonic peaks and the original odd harmonics $\Delta \omega = \Omega-O_{q}$ is then expressed as $\Delta\omega = (\frac{T_{0}}{2\Delta t}-1)O_{q}-\Delta S/\Delta t$. There are two terms contributed the frequency shift. The first is originated from the deviation of slit gap $\Delta t$ compared with $T_{0}/2$. It is proportional to the harmonic energy $O_{q}$. The second term is originated from the phase difference $\Delta S$ between the trajectories through two slits, and its influence is modulated by the slit gap. When driven by a multi-cycle laser field, the trajectories are symmetric in adjacent half cycle. One can easily obtain $\Delta t = T_{0}/2$ and $\Delta S = 0$, and then the harmonic peaks are located on $\Omega = O_{q}$.

\begin{figure}[!t]
	\includegraphics[width=8.6cm]{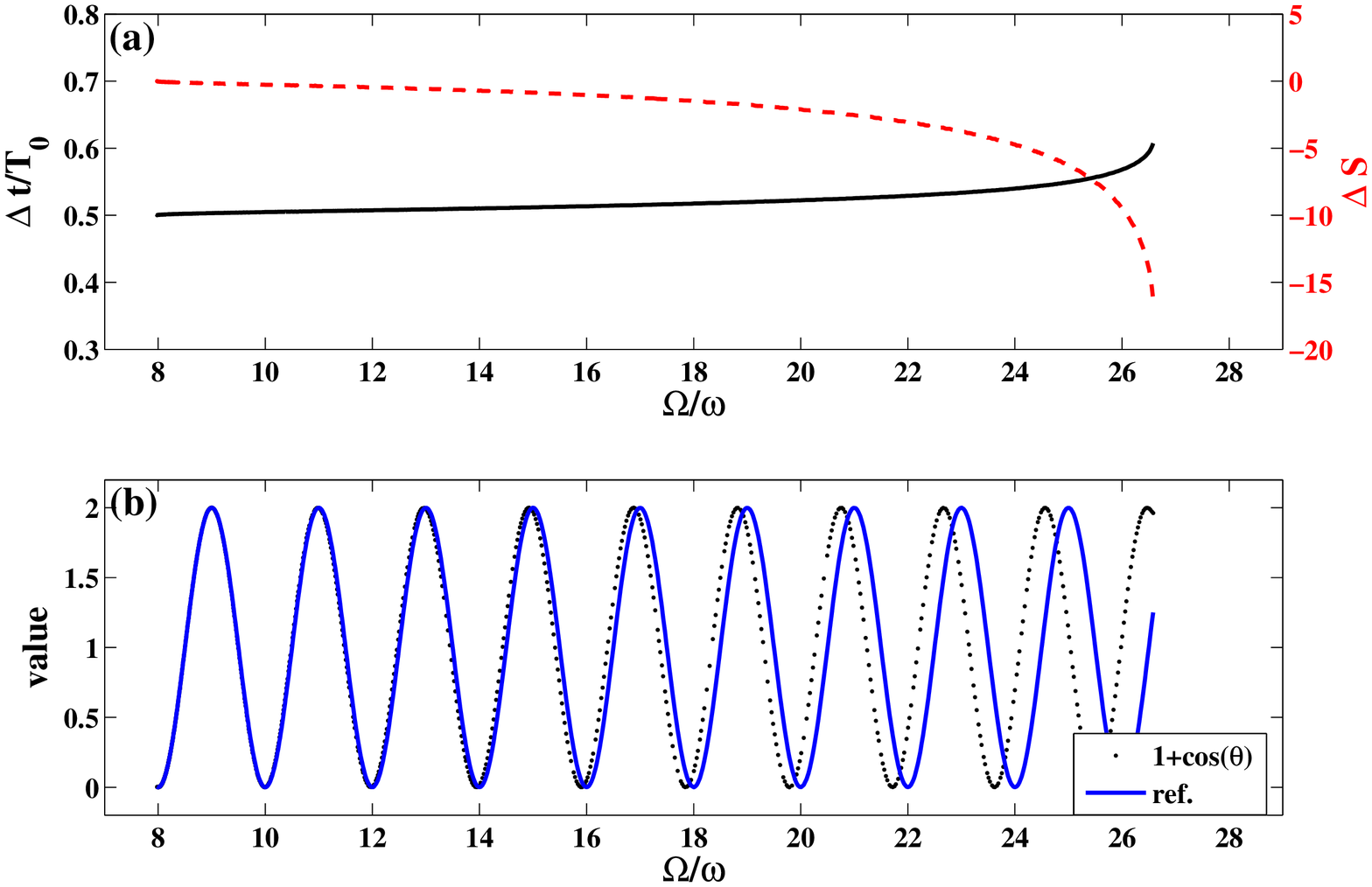}
	\caption{(a) The time delay $\Delta t$ (solid black line)  and the phase difference $\Delta S$ (dashed red line) versus photon energy. (b) Interference fringe of the two-slit interferometer (solid black line) and the reference fringe (solid blue line) with $\Delta t = T_{0}/2$ and $\Delta S = 0$. The CEP is 0.}\label{fig4}
\end{figure}

By analyzing the quantum trajectories, one can obtained the slit gap $\Delta t$ and the phase different $\Delta S$ in the two-slit interferometer. Considering the acceleration theorem and the energy conservation, the relation between the photon energy and emission time can be obtained according to the saddle point equations \cite{Vampa20151,Takuya2017,Jiang2017}:
\begin{align}\label{trajectory}
 &\int_{t_{i}}^{t}\partial_{k_x}E_{c}(-Ax(t_i)+Ax(t'))-\partial_{k_x}E_{v}(-Ax(t_i)+Ax(t'))dt' = 0 \nonumber  \\
 &E_{c}(-Ax(t_i)+Ax(t))-E_{v}(-Ax(t_i)+Ax(t)) = \Omega
\end{align}
The lines in Fig. \ref{fig3}(a), (c) shows the results predicted with Eqs. \ref{trajectory}. The blue solid lines show the photon energy as a function of emission time and the dashed black lines show the photon energy as a function of ionization time. The phase accumulated by the electron through different paths can be expressed as $S_{i} = \int_{t_{i}}^{tr_{i}}[E_{c}(-\textbf{Ax}(t_{i})+\textbf{Ax}(t'))-E_{v}(-\textbf{Ax}(t_{i})+\textbf{Ax}(t'))]dt'$, where $t_{i}$ and $tr_{i}$ are the ionization time and emission time, respectively. In Fig. 3(d), we show the trajectories ``1'' (pink solid line) and ``2'' (red solid line) in energy domain. The pink dashed line is a copy of trajectory ``1'' moved by $\Delta t$. One can see that the trajectories are modulated in adjacent half cycles, which leads to a phase difference between the two emissions. The time delay $\Delta t$ (solid red line) and the phase difference $\Delta S$ (dashed red line) between the two emissions are shown in Fig. \ref{fig4}(a). One can see that these two emissions are not separated by exactly half cycle. Instead, the time delay between these two emissions, i.e., the slit gap $\Delta t$, gradually increases with the increase of photon energy. On the contrasty, the phase difference $\Delta S$ gradually decreases.

According to Eq. \ref{peaks}, slit gaps larger than $T_{0}/2$ will give rise to a red shift while negative phase difference will give rise to blue shift comparing to the odd harmonics. With the slit gaps and the phases differences, one can obtain the interference fringe. The results for CEP $\phi=0$ are shown as dashed black lines in Fig. \ref{fig4}(b). One can see a larger red shift for the higher harmonics, which can be attributed to the bigger slit gaps [see Fig. \ref{fig4}(a)]. By changing the CEP, the two channels will move under an overall envelope, which leads to different slit gaps $\Delta t$ and phase differences $\Delta S$. Following the same procedure as we do in Fig. \ref{fig4}, the CEP dependence of the constructive interference peaks can be obtained. As shown by the stars in Fig. \ref{fig2}, the prediction of two-slit interferometer agrees very well with the SBE simulations. The slight discrepancies between a two-slit interference and simulated experiments are induced by an additional emissions that emerge with increasing the CEP (see details in Section B in supplementary material \cite{SM}).

\begin{figure}[!t]
	\includegraphics[width=8.6cm]{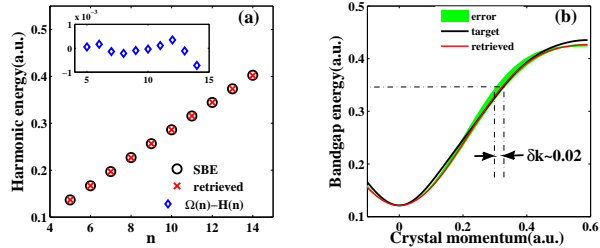}
	\caption{(a) The retrieved harmonics' frequency and the results simulated with SBE. For clarify, the difference between the SBE simulation and retrieved results are shown in the inset. (b) The target and retrieved band structure.}\label{fig5}
\end{figure}

Next, we discuss how to retrieve the band structure from the the interference fringe, i.e., the peaks $\Omega$ in the HHG spectrum. We refer the frequency of the harmonic from the simulated experiment as $H_{\phi,n}$ for a given CEP, e.g., $\phi=0$. As in \cite{Lanin}, we expand the band gap with the Fourier-series,
\begin{equation} \label{band}
\epsilon(k_{x},\textbf{c}) = \epsilon_{g}+\sum^{5}_{s=1} c_{s}cos(sk_{x}a_{x}).
\end{equation}
We assumes a known minimum band gap $\epsilon_{g}$, which can be
accurately measured with linear optical method. Then, the above equation \label{band} has 5 independent parameters $\textbf{c} = \{c_{1},c_{2},c_{3},c_{4},c_{5}\}$. Ten points from the simulated experiment are used, i.e., the circles in Fig. \ref{fig5}(a), which is well enough to determined a band structure with five independent parameters. The parameters $\textbf{c}$ can be obtained by using an self-consistent iterative method (see details in Section C of the supplementary material \cite{SM}). Our iterative method is more efficient than the best fitting algorithm, especially when the parameter space become larger. Moreover, our reconstruction is based on the temporal interferometer, without calculating the harmonic spectra as a function of laser parameter. Therefore, the band structure can be determined by a single-shot measurement. By considering that the few-cycle pulse is usually too short compare to the ultrafast processes in solids, one can ignore the modification of the band structure during the few-cycle laser pulse. Then, the single-shot measurement can be easily combined with the pump-probe scheme, which facilitate tracking the modification of band structure in real time.

Figure \ref{fig5}(a) shows the retrieved harmonics' frequency. The difference is less than $10^{-3}$ compared to the simulated experiment. Figure \ref{fig5}(b) shows the retrieved (red line) band structures obtained from the simulated experiment with $\phi=0$. One can see that the retrieved band structure well reproduces the target band (black line) with only a small difference near the maximum band energy. The deviations mainly originate from the relatively lower cutoff energy of the saddle points. As shown in Fig. \ref{fig5}(a), the highest energy is about 0.40 a.u.. In that case, the simulated results can not give an accuracy revised band data near the the maximum band energy 0.44 a.u.. We also reconstruct the band structure by using the simulated experiment with different $\phi$. The retrieved band structures are also lie very close to the target \cite{SM}. In realistic experiments, there are always some uncertainties in the laser parameters. To evaluate this influence, we consider that the fluctuations of the intensity and CEP of laser fields are $\pm5\%$ and $\pm50$ mrad, respectively. Under these conditions, the uncertainty of the reconstruction is shown in Fig. \ref{fig5}(b) with the green shadow curves. The momentum resolution amounts to $\delta k\sim0.02$ a.u..

In conclusion, we present a temporal Young's interferometer and demonstrate its applications for retrieving band structure. In our scheme, the reconstruction is based on the general relationship between the frequency shift and the modulation of the time slits. By monitoring the frequency shift of HHG in a few cycle laser field, we can directly retrieve the band structure of ZnO by a single-shot measurement. As a time-energy domain interferometric device, the temporal Young's interferometer is anticipated to possess advantageous time resolving capability. The high temporal resolution and all-optical single-shot measurement make it suitable to study matters under ambient conditions and it paves the way to track the ultrafast processes with the pump-probe approach.

\begin{acknowledgments}
This work was supported by National Natural Science Foundation (NSFC) of China (Grants No. 11627809, No. 11874165, No. 11704137, and No. 11774109), National key research and development program (2017YFE0116600).
\end{acknowledgments}

\end{document}